\documentclass[preprint,aps,a4paper, superscriptaddress,showpacs]{revtex4}
\usepackage{graphicx}

\begin{document}
\title{Coherence thresholds in models of language change and evolution: the effects of noise, dynamics 
and network of interactions}

\author{J. M. Tavares}
\affiliation{Centro de F{\'\i}sica Te{\'o}rica e Computacional and 
Departamento de F{\'\i}sica, Faculdade de Ci{\^e}ncias da Universidade de 
Lisboa, P-1649-003 Lisboa Codex, Portugal}
\affiliation{Instituto Superior de Engenharia de Lisboa\\ Rua Conselheiro 
Em\'{\i}dio Navarro, 1, P-1949-014 Lisboa, Portugal}
\author{M. M. \surname{Telo da Gama}}
\author{A. Nunes}
 \affiliation{Centro de F{\'\i}sica Te{\'o}rica e Computacional and 
Departamento de F{\'\i}sica, Faculdade de Ci{\^e}ncias da Universidade de 
Lisboa, P-1649-003 Lisboa Codex, Portugal}

\begin{abstract}

A simple model of language evolution, proposed in \cite{K_N}, is characterized by a pay-off in communicative function, 
and by an error in learning, that measures the accuracy in language acquisition. In the mean field approximation, 
this model exhibits a critical coherence threshold, i.e. a minimal accuracy in the learning process is required 
to maintain linguistic coherence. In this work, we analyse in detail the effects 
of 
different fitness based dynamics driving linguistic coherence and of the network of interactions
on the nature of the coherence threshold, by performing numerical simulations and theoretical analyses 
of generalized replicator-mutator dynamics in populations with two types of structure: fully connected 
networks 
and regular random graphs. 
We find that although the threshold of the replicator-mutator evolutionary model is robust with respect to the 
structure of the network of contacts, the coherence threshold of related fitness driven models may be strongly 
affected by this feature.

\end{abstract}


\maketitle

\section{introduction}

Statistical physics has become a powerful framework to investigate the collective behavior of individuals and is 
playing an increasingly prominent role in quantitative social sciences studies. A case in point is opinion dynamics 
{\cite{granada}} that aims at describing the emergent social behaviour, by considering models with simple rules of 
opinion formation through which 'agents' update their internal state, or opinion, through the interactions with 
other 'agents'. The interactions are typically local rules that consist in (a) following the majority or (b) random 
neighbour imitation, two simple mechanisms that have been studied for decades as models for the dynamics of Ising 
spin systems, known in the physics literature as the Glauber and Voter models, respectively {\cite{glauber,voter}}.

Traditional statistical physics models consider particles (spins, agents) interacting (i) with all the other 
particles as analytical solutions are often possible in this mean-field limit or (ii) with a number of neighbours 
located on the vertices of regular lattices in d dimensions, the topology characteristic of crystalline solids.
Recently, however, the field of complex networks {\cite{nets1,nets2}} paved the way for a better description of 
social dynamics, by providing adequate models for networks of social interactions that are neither well 
mixed as in (i) nor completely regular as in (ii). Since then, numerous studies have considered the evolution 
of opinion models on complex networks, and investigated the effects of the network topology on the model's 
dynamical behavior. In particular, novel, non-trivial behaviour has been found for the ordering dynamics of the 
zero temperature Glauber and Voter models on complex networks \cite{opin1,opin2,opin3,opin4,opin5,opin6}.
 
Language competition may be viewed as a particular case of consensus problems and as such has motivated related
studies {\cite{lang1,lang2}}. Other aspects of language dynamics include language change and evolution and language 
learning. In this context we follow the pioneering work of {\cite{K_N}} and consider an evolutionary game model where 
the errors in learning are assumed to be the major determinant for language change. This class of models for language 
change are based on the assumption that languages evolve like individuals in a population: the fittest survive and 
spread, the less fit are eliminated. The two driving forces evolution, selection and mutation (i.e.  
language transmission with a bias that favours the fittest or the dominant
language
and errors in the transmission process), are incorporated into the replicator-mutator dynamics 
equations and the time scale for change is generational.  

In the framework of evolutionary models for language dynamics, the question that arises is: how accurately 
children have to learn the language of their parents in order for the population to maintain a coherent 
language? The question was answered in a series of papers 
{\cite{K_N,K_N2,K_N3}} 
that show  
that in the strong selection limit a critical threshold, largely determined by the error rate of language 
acquisition, exists for infinite (well mixed) populations, irrespective of the number of languages in competition 
{\cite{K_N}}. 

More recently, language games such as the naming game used to model the emergence of language understood as a 
consensual lexicon {\cite{naming1,naming2}} have attracted the attention of the physics community {\cite{loreto1,loreto2}}. 
This class of models focuses on the horizontal transmission and 'creation' of language as a result of peer-to-peer 
interaction, in contrast with vertical transmission, the basic scheme of language change in models inspired by 
biological evolution {\cite{K_N}}. Apart from the time scale for change, which is no longer generational, here the 
question is to establish when the dynamics of a set of interacting agents that can choose among several options leads to 
consensus, or alternatively, when a state with several coexisting options, or language diversity, prevails. The discovery 
of critical thresholds in the naming game is arguably the most important contribution of statistical physics in a problem 
of computational linguistics.

Another class of evolutionary models of languages with analogies with the theories of population genetics was proposed by
Baxter et al {\cite{McKane}}. The model was solved in the limit of a single speaker as well as for multiple speakers in the 
mean-field approximation and (in these limits) it was shown to be related to the model of Abrams and Strogatz {\cite{AS}} 
for the extinction of languages.

In this work we follow the view on evolutionary models of language dynamics proposed by {\cite{K_N}} and consider the 
investigation of noise induced thresholds for linguistic coherence. In other words, we focus on the study (both for 
deterministic and stochastic versions) of the effectivness of the rate of learning errors, in precluding the emergence of 
linguistic consensus. In this framework the coherence threshold is the error rate of language acquisition above 
which a multi-lingual community is stable and below which there is a single dominant language. 

In the following, we extend the work of {\cite{K_N}} by analysing (i) a family of fitness driven models based on the 
replicator-mutator dynamics, that reduce to the Glauber and Voter models in the limit of neutral evolution and (ii) 
non-trivial networks of interaction. 
The results detailed below are based on the original Komarova-Nowak {\cite{K_N}} model with these generalizations, 
although the time scale for change through learning and selection is no longer generational. As in other models 
of social interaction and opinion dynamics, learning and selection occur on a shorter time scale, associated with 
'horizontal' (peer-to-peer) interaction.

The basic assumptions of our family of models are that each individual in the population is a speaker of one of two 
languages $1$ or $-1$, and that an individual may change its language through interactions with its neighbours. These 
interactions follow certain rules, where the fitnesses of the individual and of its neighbours determine the probability 
for language change in the absence of errors. In line with the usual replicator-mutator dynamics the state update that 
comes out of these rules is reversed with probability $u$, that models learning errors as the presence of noise in the 
system coupled to the dynamics. In order to assess the robustness of the coherence threshold of the replicator-mutator 
dynamics we consider models with more general fitness driven rules, that reduce to the Voter and Glauber models in the 
limit of neutral evolution and zero noise. The latter are models of spin dynamics, used to model the mechanisms of opinion 
dynamics and cultural evolution, that play a role in the evolution of languages on short time scales. 

We find that, in general, in well mixed populations dynamical noise is not effective in driving a critical coherence 
threshold. In models that reduce to the Voter and Glauber dynamics in the limit of neutral fitness the noise induced thresholds, 
separating a dominant language regime from the regime where different languages co-exist, become non-critical in the mean-field 
limit.
We derive analytical solutions for the coherence thresholds of the models in complete and regular random graphs, that are shown
to provide a very good description of the different types of cooperative behaviour of this family of models. In particular, the 
increase in robustness of the coherent linguistic regime as the number of neighbours increases is described quantitatively by the 
analytical solutions, for all dynamical models.  

Finally, we put our results in a more general context and provide a complete classification of the threshold behaviour 
of a family of fitness driven models models that includes a flipping rate, or noise uncoupled to the dynamics, instead of the 
dynamical noise of the replicator-mutator equations.

\section{Fitness driven models and dynamics}

We consider the simplest case of the model introduced in \cite{K_N} characterized by strong selection 
and two equally fit languages, with no affinity between them. Following \cite{K_N}, we consider a 
population of $N$ individuals, where each individual $i$ speaks one of two languages $\sigma_i=\pm1$ 
and define the fitness $f_i$ of $i$ as the number of its neighbors that speak the same language, 
\begin{equation}
\label{eq:fitndef}
f_i=\sum^{\prime}_{j}\delta_{\sigma_i, \sigma_j}
\end{equation}
where $\sum^{\prime}_{j}$ is a sum over the neighbors of $i$ and $\delta_{k,l}$ is $1$ if $k=l$ and $0$ 
otherwise. 
The evolution of the language follows two general rules: 
(i) the language of the fittest individuals at a given time step (generation) has a higher 
probability of being learnt by the population in the next time step (selection); (ii) in the process 
of learning there is a probability of error, i.e. a probability that the new generation learns a language 
with a lower fitness (mutation).

It was shown in \cite{K_N} that, in well mixed infinite populations, this model exhibits a critical coherence 
threshold, determined by the rate of learning errors, below which a dominant language is established and 
maintained in the population. In what follows we analyse the robustness of the linguistic coherence 
threshold when other mechanisms 
of evolution and networks of interaction are considered. 

In particular, we consider 
two generalizations of the replicator-mutator model by introducing: 
(i) more general fitness driven dynamics (including additional 
imitation/social pressure mechanisms) and (ii) populations with non-trivial interaction networks.  

We define the social fitness of an individual speaker, $F_{\pm}(i,t)$, as the total fitness of the neighbors of $i$ 
that speak language $\pm 1$ at generation $t$:
\begin{equation}
\label{Fpm}
F_{\pm}(i,t)=\sum^{\prime}_{j}f_j(t)\delta_{\sigma_j(t),\pm 1}
\end{equation}
and denote by $u$ the probability of learning errors ($0\le u \le 1$). Unless otherwise stated
the population has a fixed number $N$ of individuals. The language (of the population) at a given time 
step, $t$, is characterized by the array $\left\{\sigma_1(t),\sigma_2(t) \ldots \sigma_N(t)\right\}$. In the next 
time step the language is determined by the probability that each 
individual changes its language, through the combined effect of the dynamics and learning errors.
We have considered three fitness driven dynamical models, with learning errors, as detailed below.

\subsection{Replicator-mutator dynamics}

In this model the noise, or rate of learning errors, is incorporated in the probability of language change.
If $\sigma_{i}(t-1)=1$ the probability for language change (i.e., the probability that 
the outcome of the update rule is $\sigma_{i}(t)=-1$) is given by {\cite{K_N}}, 
\begin{equation}
\label{p1m1nk}
P_{1\rightarrow -1}=\frac{(1-u)F_{-}(i,t-1)+uF_{+}(i,t-1)}
{F_{-}(i,t-1)+F_{+}(i,t-1)}.
\end{equation}    
while if $\sigma_{i}(t-1)=-1$ the probability for language change (i.e., the probability that 
the outcome of the update rule is $\sigma_{i}(t)=1$) is given by,
\begin{equation}
\label{pm11nk}
P_{-1\rightarrow 1}=\frac{(1-u)F_{+}(i,t-1)+uF_{-}(i,t-1)}{F_{-}(i,t-1)+F_{+}(i,t-1)}.
\end{equation} 

\subsection{Fitness driven Voter dynamics}

The update rule for this model is inspired in the simplest opinion/imitation dynamics model, 
the Voter model: a speaker changes language if its fitness is lower than a randomly chosen 
neighbour that speaks a different language. In addition, with probability $u$ the outcome of 
the dynamical rule is reversed. The update rule for a speaker $i$ at time $t$ is the following:  
(i) choose one neighbor, $j$, at random; (ii) if $f_j(t-1)> f_i(t-1)$ then $\sigma_i(t)=\sigma_j(t-1)$; 
(iii) if $f_j(t-1)\le f_i(t-1)$ then $\sigma_i(t)=\sigma_i(t-1)$; (iv) reverse the outcome of the dynamical 
rule with probability $u$.

\subsection{Fitness driven Glauber dynamics}

The update rule for this model is inspired in the Glauber dynamics of spin systems, that mimicks 
the effect of social pressure in opinion dynamics. In this model each individual tends to adopt the 
fittest (dominant) language in its neighborhood. The update rule for a speaker $i$ at time $t$ is the following:
(i) if $F_{+}(i,t-1) > F_{-}(i,t-1)$ then $\sigma_{i}(t)=1$; 
(ii) if $F_{+}(i,t-1) < F_{-}(i,t-1)$ then $\sigma_{i}(t)=-1$;
(iii) if $F_{+}(i,t-1) = F_{-}(i,t-1)$ then $\sigma_{i}(t)=\sigma_{i}(t-1)$;
(iv) reverse the outcome of the dynamical rule with probability $u$.

\section{Simulations for well mixed populations and for regular random graphs}

The models described above were first simulated on complete graphs or fully connected networks, i.e., where all 
individuals are neighbors of each other. On these networks the fitness of individuals speaking the same 
language is identical, in each time step (generation): the fitness of individuals speaking $+1$($-1$) is 
$N_1-1$ ($N-N_1-1$), where $N_1$ is the total number of speakers of $1$. 
  
We start the simulations from a fully ordered system, i.e., at $t=0$ all individuals speak language $+1$ (say). 
The language of the next generation is determined according to the rules described above for each model.
The language of the population evolves through a large number of generations (5000) and the mean value of 
$x=\frac{N_1}{N}$ is calculated for each value of error rate in learning, $u$.  

In figure 1 we plot the results of simulations of the three dynamics, for different population sizes, $N=100$, 
$N=1000$ and $N=10000$. Although finite size effects are visible for the smaller systems they are negligible for 
populations of thousands. While we find a critical coherence threshold, at $u=1/4$, reproducing the results of 
{\cite{K_N}} for the Komorova-Nowak model, the results for the Voter and Glauber fitness driven models are quite 
different: the threshold is shifted to the value of the noise that completely overrides
the dynamics, $u=1/2$, and the fraction of speakers of the dominant language 
approaches $x=1/2$
linearly, at the 
threshold, revealing its non critical nature.

In order to investigate the effects of the network of interactions on the linguistic coherence threshold 
we simulated the same models on regular random graphs (RRG), where analytical results may also be obtained.
In RRG networks $N$ nodes are linked at random to a fixed number of neighbors, $k$, without double 
and self links. The models were simulated on two of these networks, for small ($k=4$) 
and large ($k=20$) degree. The simulations, for $N=10^3$ and $N=10^4$, start (as before) in the ordered state 
where all individuals speak $+1$. For each value of $u$, the language evolves through 10000 generations at the 
end of which the average fraction of speakers of the dominant language is computed. The results for the three 
models are plotted in figures 2,3 and 4. 

Note that the transition on RRG exibhits a critical threshold for all models. The value 
of $u$ at threshold, $u_{th}$, increases with the number of neighbours and approaches the MF values 
($\frac {1}{4}$ for the replicator mutator and $\frac{1}{2}$ for the voter and Glauber dynamics) as the 
number of neighbours tends to infinity. Above threshold, $u>u_{th}$, the equilibrium value of $x$ corresponds 
to the coexistence of the two languages, $x=\frac{1}{2}$. 

\section{
Analysis of the mean field equations}

In order to shed light on these results we proceed to calculate the equilibrium values of $N_1$ analytically, 
in the infinite population limit.
Let $x$ be the fraction of speakers of language $1$ ($x\equiv N_1/N$). 
The evolution of $x$ is given by,
\begin{equation}
\label{eq:evolxtime}
\dot{x}= - x P_{1\rightarrow -1} + (1-x) P_{-1\rightarrow 1}
\end{equation}
where $P_{1\rightarrow -1}$ and $P_{-1\rightarrow 1}$ are the rates of change of the two competing languages. 

In well mixed 
populations $P_{1\rightarrow -1}$ and $P_{-1\rightarrow 1}$ depend only on $x$ and 
can be computed exactly for the three models. 

On RRG networks these probabilities are calculated using the following (mean-field) assumptions:
(i) each of the $k$ neighbours of any site are linked to $(k -1)$ second neighbors, with 
no loops (uncorrelated links);  
(ii) the probability that the language spoken at a given site is $+1$ is the average density 
of speakers of that language, $x$ (uncorrelated densities).
Within this mean-field approximation, the fitness of each node is a random variable that results 
from the sum of $k$ independent and identical binomial variables. The calculation of the transition probabilities $P_{1\rightarrow -1}$ and $P_{-1\rightarrow 1}$ depends on the specific dynamics and 
proceeds in a straightforward fashion.

Given the symmetry of the models the probabilities (\ref{eq:evolxtime}) may be written as,
\begin{eqnarray}
\label{p1m1}
P_{1\rightarrow -1} & = & (1-u)Q(x)+u(1-Q(x)) \\
\label{pm11}
P_{-1\rightarrow 1} & = & (1-u)Q(1-x)+u(1-Q(1-x)),
\end{eqnarray} 
where $Q(x)$ is a function that depends on the network of contacts and on the dynamics. 
Substituting (\ref{p1m1},\ref{pm11}), (\ref{eq:evolxtime}) becomes
\begin{equation}
\label{xdot}
\dot{x}=(1-2u)\left( -xQ(x)+(1-x)Q(1-x)\right)+u(1-2x).
\end{equation}
   
In what follows we discuss the meaning of $Q(x)$ for each dynamical model and calculate it for each network of contacts.
 
\subsection{Replicator-mutator dynamics}

We find by inspection of (\ref{p1m1nk},\ref{pm11nk}) and (\ref{p1m1},\ref{pm11}) 
that for the replicator-mutator model, $Q(x)$ ($Q(1-x)$) is the normalized value of $F_{-}$ ($F_{+}$) 
in the neighbourhood of a speaker of language $1$ ($-1$),   
\begin{equation}
\label{Qrm}
Q(x)=\frac{F_{-}}{F_{-}+F_{+}}.
\end{equation}

In well mixed populations, $F_{-}$ and $F{+}$ take their mean values, and $Q(x)$ becomes,
\begin{equation}
\label{Qrmmf}   
Q(x)=\frac{(1-x)^2}{(1-x)^2+x^2}.
\end{equation} 
Substituting (\ref{Qrmmf}) into (\ref{xdot}) yields the evolution equation,
\begin{equation}
\label{evolxtimemf}
\dot{x}=\frac{\left(-x(1-x)+u\right)\left(1-2x\right)}{x^2+(1-x)^2},
\end{equation} 
and the (stable) fixed points, $x^*$, are solutions of $\dot{x}=0$. It is 
straightforward to show that 
\begin{equation}
\label{eqnk}
x^*=\left\{ \begin{array}{lll}
\frac{1}{2} & \textrm{if $u>\frac{1}{4}$}
\\
\\
\frac{1}{2}\left(1\pm\sqrt{1-4u}\right) & \textrm{if $u\le \frac{1}{4}$}
\end{array}\right.
\end{equation}
confirming that $u_{th}=\frac{1}{4}$ is the threshold for linguistic 
coherence. 
Furthermore, this threshold is critical since the derivative of $x^*$ w.r.t. $u$ diverges there. The 
function (\ref{eqnk}) is plotted, for $x^*\ge \frac{1}{2}$, in figure 1, and excellent agreement is 
found between the analytical solution and the simulation results for large systems.

The transition probabilities of the replicator-mutator model on RRG are calculated by determining 
the average value of the total fitnesses $F_{+}(i)$ and $F_{-}(i)$ in the neighborhood of a given node. 
Let us consider a node $i$ that 
speaks $+1$ with $n$ neighbours that speak also $+1$. The average fitness of one of these neighbours is 
$1+(k-1)x$ and that of the neighbours speaking $-1$ is $(k-1)(1-x)$. The average values of $F_{+}(i)$ 
and $F_{-}(i)$ are then,
\begin{eqnarray}
\label{Fmnk}
F_{-}=(k -n)(k-1)(1-x),
\\
\label{Fpnk}
F_{+}=n(1+(k  -1)x).
\end{eqnarray}
The number of neighbours of $i$ speaking the same language, $n$, is a random variable that results from 
the sum of $k$ random binomial variables, each one taking the value $1$ with probability $x$ and $0$ with 
probability $(1-x)$. The function $Q(x)$ in (\ref{Qrm}) is then,
\begin{equation}
\label{Qrmrrg}
Q(x)=\sum_{n=0}^{k} 
B(k,n)x^n (1-x)^{k -n} 
\frac{F_{-}}
{F_{-}+F_{+}},
\end{equation}
where $F_{-}$ and $F_{+}$ are given by (\ref{Fmnk},\ref{Fpnk}) and $B(i,j)$ is the binomial coefficient, 
\begin{equation}
\label{bincoef}
B(i,j)=\frac{i!}{j!(i-j)!}.
\end{equation}
The evolution equation is obtained by substituting (\ref{Qrmrrg}) in (\ref{xdot}). 
The stable fixed points as a function of the noise parameter are ploted in figure 2 for 
$k=4$ and $k=20$, respectively. For regular random graphs with $k =20$ the agreement between the simulation 
and the analytic results is almost quantitative, for populations of the order of a few thousand.

\subsection{Fitness driven Voter dynamics}

In the fitness driven Voter dynamics an 
individual that speaks language $1$ changes to language $-1$: 
with probability $(1-u)$ if the neighbour chosen at random speaks $-1$ and has a higher fitness; 
with probability $u$ if the neighbour chosen at random speaks $1$ or has a lower fitness.  
Thus, $Q(x)$ ($Q(1-x)$) is the probability to find a neighbour that speaks $-1$ ($1$) 
and has a higher fitness.
In the mean field approximation, 
$Q(x)$ is
\begin{equation}
\label{eq:Qvdrnet}
Q(x)=(1-x)H(x),
\end{equation}
the product of the probability $(1-x)$ of  
finding a neighbor that speaks $-1$ and the probability $H(x)$
that a speaker of $-1$ has higher fitness than a speaker of $1$. 

In well mixed populations, the probability that a speaker $-1$ has a higher fitness is $1$ ($0$) for 
$x<\frac{1}{2}$ ($x>\frac{1}{2}$) implying that,
\begin{equation}
\label{q1vm}
Q(x)=(1-x) \Theta(1-2x),
\end{equation}
where $\Theta(z)$ is the step function: $\Theta(z)=1$ if $z>0$ and $\Theta(z)=0$ if $z\le0$. 
The dynamical equation is obtained by substituting (\ref{q1vm}) into (\ref{xdot}) and has stable 
fixed point solutions, $\dot{x}=0$, given by
\begin{equation}
\label{eqvm}
x^*=\left\{ \begin{array}{lll}
\frac{1}{2} & \textrm{if $u\ge\frac{1}{2}$}
\\
\\
\frac{1}{2}\left(1\pm(2\alpha-\sqrt{1+4\alpha^2})\right) 
& \textrm{if $u< \frac{1}{2}$}
\end{array} \right .,
\end{equation}
where $\alpha=\frac{u}{1-2u}$. 
Again, the rate of learning errors $u^{th}=\frac{1}{2}$ defines two 
regimes: 
for $u<u^{th}$ a dominant language is established and maintained while for $u\ge u^{th}$ there is coexistence of 
the two equally probable languages.   
Note, however, that $u^{th}=1/2$ is a trivial threshold in the sense that for this level
of noise the evolution is totally random, while for higher levels of noise the evolution rules
actually hinder linguistic coherence.
This trivial threshold is non critical since, 
\begin{equation}
\label{critthrvm}
\lim_{u\to {\frac{1}{2}}^{-}} \frac{dx^*}{du}=\pm\frac{1}{2},
\end{equation} 
the derivative at threshold is finite. The function (\ref{eqvm}) is plotted, for 
$x^*\ge \frac{1}{2}$, in figure 1, and apart from the finite size effects mentioned previously, quantitative 
agreement is found between the analytical solution and the simulation results.

To calculate $H(x)$ for RRG, 
let us consider a node $i$ with $\sigma_i=+1$, and a neighbor $j$ with $\sigma_j=-1$.  
Using the definition of fitness and the rules of the Voter dynamics we can compute $H(x)$ as the probability 
that $j$ has a number $m$ of neighbours speaking $-1$ that is larger than the number $n$ of neighbours of $i$ 
speaking $+1$,   
\begin{eqnarray}
\nonumber
H(x)=\sum_{n=0}^{k -2}
B(k -1,n) x^n (1-x)^{k -1-n} \times
\\
\label{eq:Hvotd}
\times \sum_{m=n+1}^{k -1}
B(k-1,m) (1-x)^m x^{k -1-m}.
\end{eqnarray}
The stable fixed points are calculated using (\ref{xdot}) with $Q(x)$ given by 
(\ref{eq:Qvdrnet},\ref{eq:Hvotd}) and solving for $\dot x =0$. In general 
($x \ne \frac{1}{2}$) they are more easily written in terms of $u(x^*)$, 
\begin{equation}
\label{fixpvd2}
u(x^*)=\left\{ \begin{array}{llll}
\frac{V(x^*)}{2x^*-1+V(x^*)} & \textrm{if $x^*>\frac{1}{2}$}
\\ 
\\
\frac{V(1-x^*)}{1-2x^*+V(1-x^*)} & \textrm{if $x^* <\frac{1}{2}$}
\end{array}\right .
\end{equation}
where $V(x)=x(1-x)(H(1-x)-H(x))$. $u(x^*)$ given by (\ref{fixpvd2}) is plotted in figure 3 for 
$x>\frac{1}{2}$. 
The calculated $x^*(u)$ is in line with the results of the 
simulation and for 
$k =20$ the agreement becomes nearly quantitative. 

\subsection{Fitness driven Glauber dynamics}

In the fitness driven Glauber dynamics, $Q(x)$ is the probability that a speaker $i$ of $+1$ has 
total fitness satisfying $F_{-}(i)>F_+(i)$.

In well mixed populations, $F_->F_+$ iff $x<\frac{1}{2}$ and therefore, 
\begin{equation}
\label{qxgd}
Q(x)=\Theta(1-2x).
\end{equation}
Substituting in (\ref{xdot}) and solving for the stable fixed points we obtain   
\begin{equation}
\label{fixpgd}
x^*=\left\{ \begin{array}{lll}
\frac{1}{2}& \textrm{if $u\ge\frac{1}{2}$}
\\
\\
\frac{1}{2}\pm(\frac{1}{2}-u)& \textrm{if $u <\frac{1}{2}$}
\end{array}\right.
\end{equation}

The rate of learning errors $u^{th}=\frac{1}{2}$ is again a trivial threshold that separates
two regimes as in the Voter driven model.
Also as in the Voter model the threshold is non-critical since, the derivative of $u$ at threshold is finite and 
the fraction of speakers of the dominant language approaches 
the threshold linearly. The function (\ref{fixpgd}) is plotted, for 
$x^*\ge \frac{1}{2}$, in figure 1, and excellent agreement is found between the analytical solution and the 
simulation results, for large systems.

In order to calculate $Q(x)$ for RRG, we consider in the neighborhood of a given node $i$ that is a $+1$ speaker:  
(i) $n$ nearest neighbors that speak $+1$; (ii) $n_1$ next nearest neighbours that speak $+1$ 
and share with $i$ a nearest neighbour that speaks $+1$; (iii) $n_2$ next nearest neighbours  
that speak $-1$ and share with $i$ a nearest neighbour  that speaks $-1$.
Then the total fitnesses are simply given by $F_{-}(i)=n_2$ and $F_{+}(i)=n+n_1$ and $Q(x)$ is the 
probability that $n_2 >n+n_1$, 
\begin{eqnarray}
\nonumber
Q(x)=\sum_{n=0}^{k}
B(k,n) x^n (1-x)^{k -n} 
\sum_{n_1=0}^{n(k -1)}
B(n(k -1),n_1) x^n_1(1-x)^{n(k -1)-n_1} \times
\\
\label{gxgd}
\times
\sum_{n_2=n+n_1+1}^{(k-n)(k -1)}
B((k-n)(k -1),n_2)
(1-x)^{n_2} x^{(k-n)(k -1)-n_2}
\end{eqnarray}
The fixed points are calculated using (\ref{xdot}) and solving for $\dot x=0$.
The function $u(x^*)$ obtained is of the form (\ref{fixpvd2}) 
with $V(x)=-xQ(x)+(1-x)Q(1-x)$ and  
$Q(x)$ given by (\ref{gxgd}). $u(x^*)$ is plotted in figure 4 
for $x>\frac{1}{2}$ and is found to be in line with 
the results of the simulations. For $k =20$ the agreement is almost quantitative. 

Again, the transition exibhits a critical threshold and the value of $u_{th}$ also increases with the 
number of nearest neighbours on the network approaching the MF value, $u_{th}=1/2$, as this number 
approaches infinity. Above threshold, $u>u_{th}$, 
the stable fixed point corresponds to the coexistence of the two languages, $x=\frac{1}{2}$.

\subsection{Coherence thresholds for social fitness driven evolution}

For all the models considered above, the mean field description of the dynamics is given by
equation (\ref{xdot}) which is of the form
\begin{equation}
\dot x = (1 - 2u) g(x) + u (1 -2x)
\label{eqgen}
\end{equation}
with $g(1/2)= 0$ and $g(x)=-g(1-x)$. For (\ref{eqgen}) to describe the 
mean field dynamics of an evolution process that selects for the 
dominant variant of two languages or species with the same intrinsic fitness,
the additional assumptions are that
$g(0)=0$ and $g(x)<0$ for $0<x<1/2$, so that
the coherent states $x=0$ and $x=1$ are the only stable solutions
in the absence of noise.
The threshold behaviour of this type of models is
easily understood if we consider the related family
\begin{equation}
\dot x = \tilde g(x) + u (1 -2x)
\label{eqdrift}
\end{equation}
where $\tilde g (x)$ has the same symmetry properties as $g(x)$ and
noise and dynamics are uncoupled, so that $u$ represents a constant
rate of random flipping independent of the evolution rules.
Indeed, (\ref{eqgen}) can be brought to the form (\ref{eqdrift}) 
with $\tilde g(x) = g(x)(1-2x)/(1-2x-2g(x))$ through a
smooth rescaling of time, provided that $g(x)$ is smooth.
From equation (\ref{eqdrift}), the curve $u(x)$ that relates the rate
of random flipping $u$ with the corresponding equilibrium density $x$ is
\begin{equation}
u(x)=\frac{\tilde g(x)}{2x-1}.
\label{eqcurve}
\end{equation}
If we assume that $\tilde g(x)$ is smooth,
then given the symmetry 
\begin{equation}
\tilde g(x) = \tilde g'(1/2) (x-1/2) + {\cal O} ((x-1/2)^3)
\end{equation}
and therefore
\begin{equation}
u^{th}= u(1/2) = \tilde g'(1/2)/2, \ \ \ \ \frac{du}{dx}^{th}=u'(1/2)=0.
\label{threshdrift}
\end{equation}
This means that models (\ref{eqdrift}), (\ref{eqgen}) always exhibit critical
coherence thresholds when $\tilde g$, $g$ are smooth. 
The values of the critical thresholds found in this section are particular cases of
equation (\ref{threshdrift}), which for model (\ref{eqgen}) reads
\begin{equation}
u^{th}= u(1/2) = \frac{1}{2} \frac{g'(1/2)}{1+g'(1/2)}, \ \ \ \ \frac{du}{dx}^{th}=u'(1/2)=0.
\label{threshgen}
\end{equation}
Equation (\ref{eqcurve}) also shows that whenever $\tilde g(x)$ is a step function
with a discontinuity at $x=1/2$, then model (\ref{eqdrift}) has no coherence threshold:
a dominant language persists for arbitrarily large levels of noise, because the 
right and left limits
$u(1/2^+)$ and $u(1/2^-)$ are both infinite. 
The behaviour of model (\ref{eqgen}) when $g(x)$ is a step function with a discontinuity
at $x=1/2$, as for the fitness driven Voter and Glauber dynamics on the complete
graph, may be obtained directly from the analogue of equation (\ref{eqcurve}) for
model (\ref{eqgen})  
\begin{equation}
u(x)=\frac{g(x)}{2x -1 + 2g(x)}.
\label{eqcurve2}
\end{equation}
Then
\begin{equation}
u^{th}= u(1/2^+)=u(1/2^-)=1/2
\end{equation}
independently of $g$, and 
\begin{equation}
\frac{du}{dx}^{th}=u'(1/2^{+,-})= - \frac{1}{2} \frac{1}{g(1/2^{+,-})},
\end{equation}
which is always bounded away from zero. Therefore, this class of models will
exhibit a non critical trivial threshold at $u=1/2$.

To summarize, the
family of models described at the mean field level by
equations (\ref{eqgen}), (\ref{eqdrift}) 
exhibits three types of threshold behaviour.
Models with non-smooth density dependence transition rates $g(x)$ and 
dynamically coupled noise exhibit a trivial non-critical threshold at the value of $u=1/2$ 
for which noise completely overrides the dynamics. 
Models with non-smooth density dependence transition rates $\tilde g(x)$ 
and dynamically uncoupled noise do not exhibit a noise induced threshold, i.e. there 
is always a dominant language irrespective of the level of noise.
Finally, in the generic case of models with smooth transition rates 
$g(x)$, $\tilde g(x)$ there is a critical 
threshold, below which language coherence may be established and maintained.

\section{Conclusions}

We have considered different models for the evolution in the presence of noise
of two languages with the same intrinsic fitness that compete through the
selective advantage of the language that is perceived by each individual as
the dominant language. The language spoken by each speaker has for that speaker
a social fitness given by the number of its neighbours that share the same language,
and the dynamics driven by evolution rules based on this fitness measure 
will depend also on the interaction network of the population.
Starting from a state where all individuals speak the
same language, mutations or transmission errors act as noise terms that
favour the balance of the number of speakers 
of each language, while selection according to 
social fitness drives linguistic coherence. The coherence threshold is the
level of noise or mutation rate above which the system evolves to a state
where both languages are equally frequent.

From simulations of these models on fully connected networks and on regular
random graphs,
we found that the critical threshold for the replicator-mutator model \cite{K_N} is robust 
with respect to the network structure, but that the coherence thresholds of related models are strongly affected by this feature.
In particular, we have found that models with 
social fitness driven dynamics inspired by the 
Voter and Glauber models, two of the simplest models for spin dynamics used in opinion dynamics 
and cultural evolution studies, exhibit different linguistic coherence threshold behaviour, depending on the network of interactions. On a regular random graph, these models 
have a critical coherence threshold, while on the fully connected network 
a dominant language persists up to the level of noise for which the evolution rules
are totally random.

We have obtained analytical mean-field solutions for the coherence thresholds 
on the fully connected network and on regular random networks that are in agreement with the results of the simulations for the three 
models, providing a quantitative description of the behaviour of the 
different microscopic rules. We have shown that the noise threshold behaviours of these
models and, more generally, of evolution processes that select for the 
dominant variant of two languages or species with the same intrinsic fitness,
can be understood as well in terms of a simple mean field analysis.

\section{Acknowledgements}

Financial support from the Foundation of the University of Lisbon 
and the Portuguese Foundation for Science and 
Technology (FCT) under contracts POCI/FIS/55592/2004
and POCTI/ISFL/2/618 is gratefully 
acknowledged.

\newpage

\section*{Figures}

\begin{figure}[!htb]
\begin{center}
\includegraphics[width=12cm]{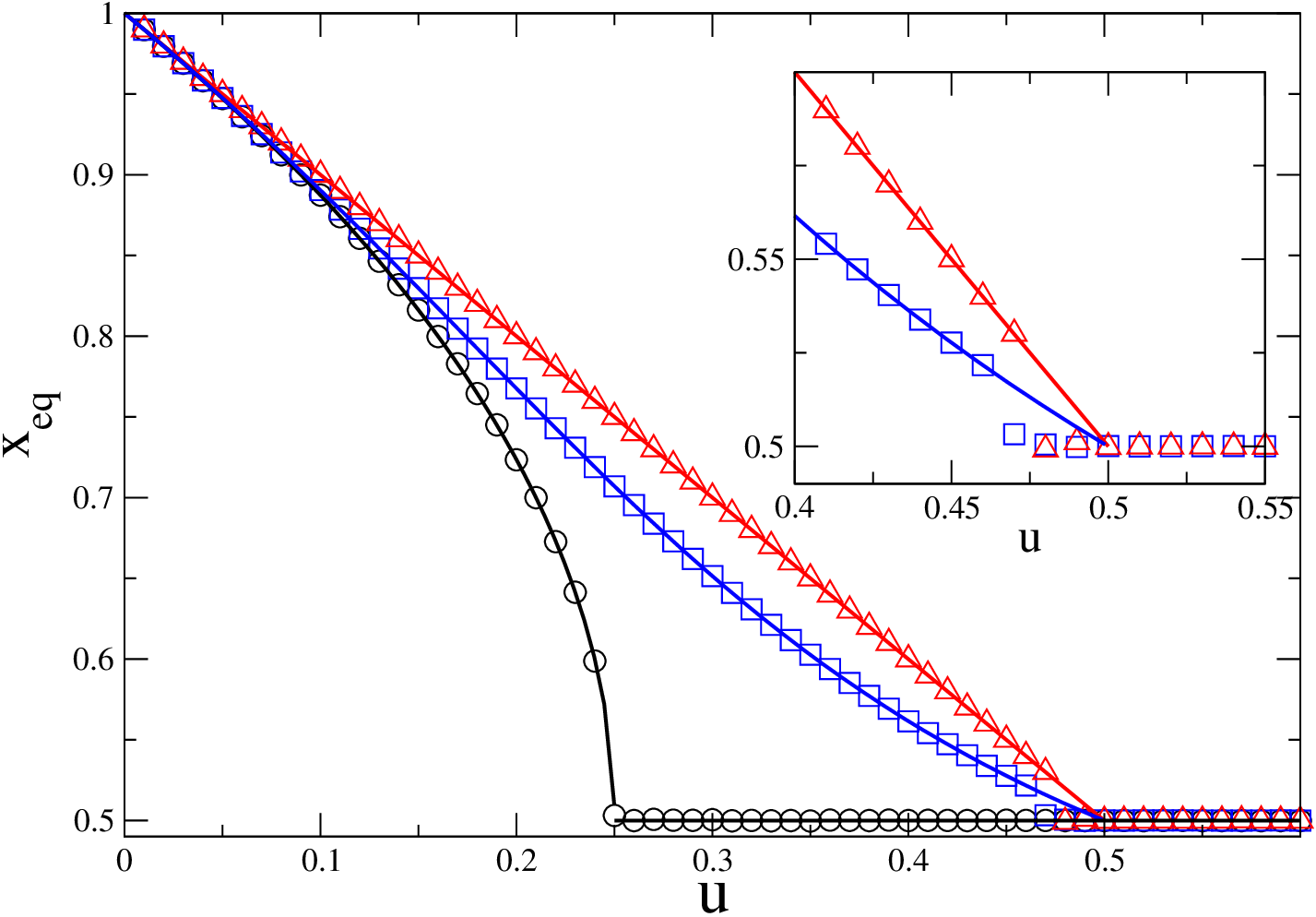}
\caption{
Symbols: mean fraction of speakers of language 1 ($x$) during 5000 generations for different values 
of $u$ and a population $N=10^4$ (simulation results in a fully connected 
 network, for an initial condition $x=1$). Circles: Replicator-mutator
dynamics; 
Triangles: fitness driven Glauber dynamics; squares: fitness driven voter dynamics. 
Lines: fixed points  $x^*$ 
from (\ref{eqnk}, \ref{eqvm}, \ref{fixpgd}) for $x\ge \frac{1}{2}$.}
\label{fig:threshmf}
\end{center}
\end{figure}

\begin{figure}[!htb]
\begin{center}
\includegraphics[width=12cm]{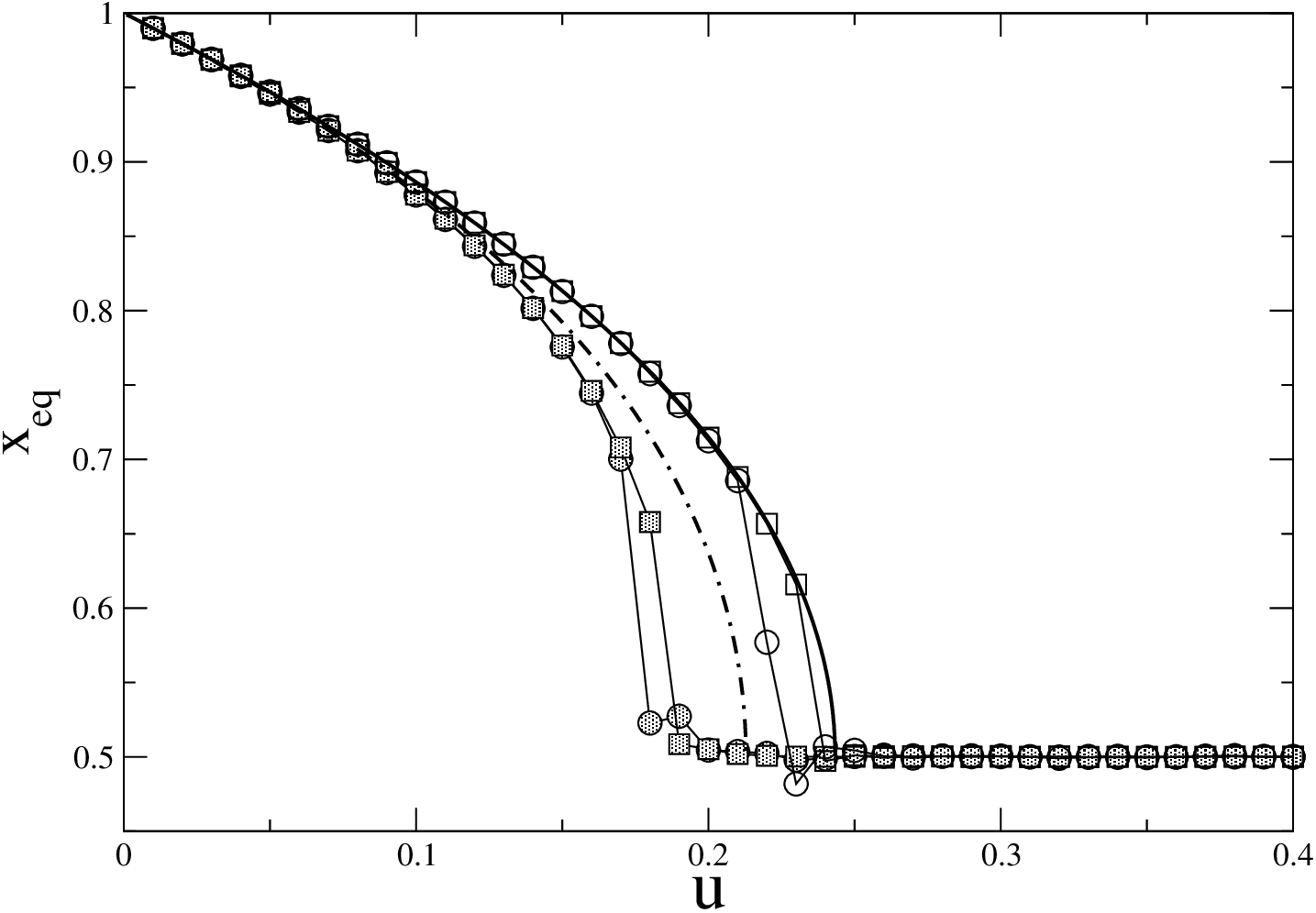}
\caption{
Symbols: mean fraction of speakers of language 1 as a function of $u$ 
for the Replicator-mutator dynamics from 
simulations (10000 generations for an initial condition $x=1$.) in a random regular lattice with degree $k$; 
full symbols: $k=4$; open symbols: $k=20$; circles: $N=10^3$; squares: $N=10^4$.  
Lines: fixed points $x^*$ from  
(\ref{xdot}, \ref{Qrmrrg}) for $x\ge \frac{1}{2}$; full line: $k=20$; dot-dashed line: $k=4$.}
\label{fig:threshrmrrg}
\end{center}
\end{figure}

\begin{figure}[!htb]
\begin{center}
\includegraphics[width=12cm]{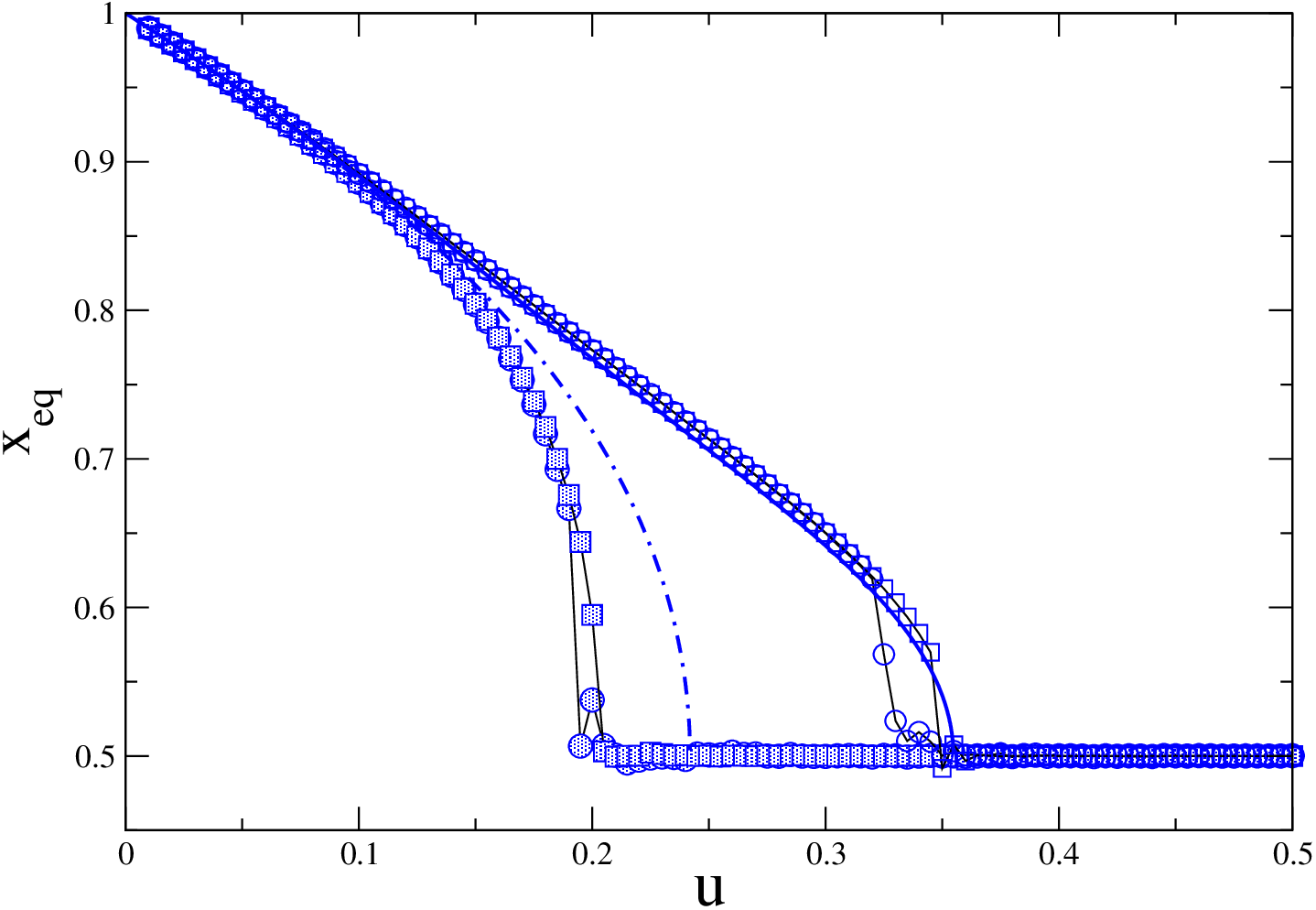}
\caption{
The same as in figure \ref{fig:threshrmrrg} but for the fitness driven voter dynamics.
The fixed points $x^*$ were calculated using   
(\ref{xdot}, \ref{eq:Hvotd},\ref{eq:Qvdrnet}) for $x\ge \frac{1}{2}$.}
\label{fig:threshvdrrg}
\end{center}
\end{figure}

\begin{figure}[!htb]
\begin{center}
\includegraphics[width=12cm]{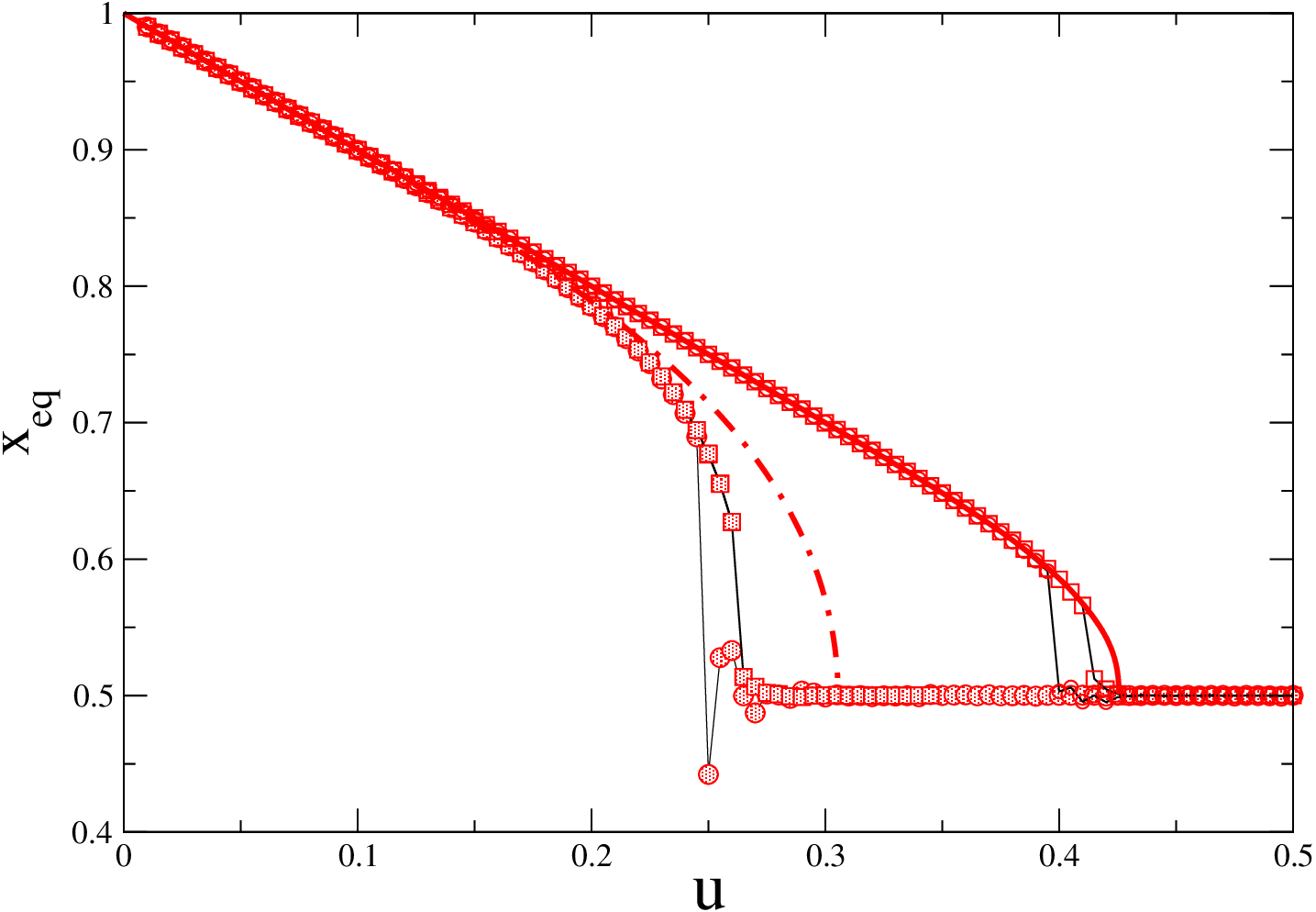}
\caption{
The same as in figure \ref{fig:threshrmrrg} but for the Glauber driven voter dynamics.
The fixed points $x^*$ were calculated using   
(\ref{xdot}, \ref{gxgd}) for $x\ge \frac{1}{2}$.}
\label{fig:threshgdrrg}
\end{center}
\end{figure}

\end{document}